\documentclass[aps,pra,superscriptaddress,twocolumn,showpacs]{revtex4}
\usepackage{amsmath}
\usepackage{amssymb}
\usepackage{bm}
\usepackage{epsfig}
\usepackage{graphicx}
\usepackage{color}

\begin{document}

\title{Spin evolution of cold atomic gases in SU(2)$\otimes $U(1) fields}
\author{I. V. Tokatly}
\affiliation{Nano-bio Spectroscopy group and ETSF Scientific Development Centre, Dpto.
F\'isica de Materiales, Universidad del Pa\'is Vasco, Centro de F\'isica de
Materiale CSIC-UPV/EHU-MPC, E-20018 San Sebastian, Spain }
\affiliation{IKERBASQUE Basque Foundation for Science, Bilbao, Spain}
\author{E. Ya. Sherman}
\affiliation{Department of Physical Chemistry, Universidad del Pa\'is Vasco UPV-EHU,
48080 Bilbao, Spain}
\affiliation{IKERBASQUE Basque Foundation for Science, Bilbao, Spain}

\begin{abstract}
We consider response function and spin evolution in spin-orbit coupled cold atomic
gases in a synthetic gauge magnetic field influencing solely the orbital
motion of atoms. We demonstrate that various regimes of spin-orbit
coupling strength, magnetic field, and disorder can be treated within
a single approach based on the representation of atomic motion in terms of auxiliary 
collective classical trajectories. Our approach allows for a unified description of fermionic
and bosonic gases. 
\end{abstract}

\pacs{03.75.Ss, 05.30.Fk, 67.85.−d}

\date{\today}
\maketitle

\section{Introduction}

Recent advances in the experimental physics of cold atomic gases led to the
observation of new regimes of their spin dynamics, both for bosons 
\cite{Stanescu,Lin} and fermions \cite{Sommer,Liu,Wang,Cheuk,Huang16}. One of the most
interesting features of these systems is a synthetic optically produced
(pseudo)spin-orbit coupling, playing the critical role there. The qualitative
effects of this coupling seen in cold matter are strongly different from the
effects observed in solids \cite{Winkler03,Zutic04,Dyakonov08,Wu10}.
Recently, in became possible to realize either optically \cite%
{Lin09} or by a mechanical rotation \cite{Cooper} synthetic magnetic fields
influencing the orbital motion of cold atomic gases. Comprehensive reviews of the 
field can be found in Ref. [\onlinecite{reviews}].

A spin-orbit coupled three-dimensional system of particles with pseudospin 1/2 can be described by Hamiltonian \cite{note1}: 
\begin{eqnarray} \nonumber
H&=&\int d^{3}r\frac{1}{2m}\Psi ^{\dagger }\left( -i\partial _{j}-A_{j}-%
\mathcal{A}_{j}^{\mathrm{\rm so}}\right) ^{2}\Psi + {\Psi ^{\dagger }\mathcal{A}^0\Psi} \\
 &+& W\left[ \Psi ^{\dagger },\Psi %
\right],  
\label{Hamiltonian}
\end{eqnarray}%
where $m$ is the particle mass, $\Psi ^{\dagger }$ and $\Psi $ are the
particle-related two-component spinor field operators, and $W\left[ \Psi ^{\dagger },\Psi \right] $ is
determined by the spin-independent external potential and interaction between the particles.
The spin-independent components of the gauge field $A_{j}$  are due to the U(1) synthetic
magnetic field {${\bf B}=\nabla\times{\bf A}$}. The spin-orbit coupling is represented here by generally a
non-Abelian {SU(2)} gauge field $\left( \mathcal{A}^{0},\mathcal{A}_{1}^{\rm so},%
\mathcal{A}_{2}^{\rm so},\mathcal{A}_{3}^{\rm so}\right) ,$ where each component is 
a $2\times 2$ matrix, $\mathcal{A}^{0}$ component corresponds to the Zeeman
coupling, and spatial components describe the spin-orbit coupling. Thus, we
consider a cold matter in a {background gauge} field with the SU(2)$\otimes $U(1)-symmetry.
This general approach to the spin-orbit coupling in terms of a SU(2)-field
leads to a deep understanding of the spin-related properties of condensed
matter \cite{Mineev92,Frolich93,Aleiner01,Levitov03,Tokatly08,Leurs08,Raimondi}. Here,
the vector potential for the linear in the particle momentum spin-orbit
coupling has the form:%
\begin{equation}
\mathcal{A}_{j}^{\mathrm{\rm so}}=\frac{1}{2}q_{j}\left( {\mathbf{h}}^{\left[ j%
\right] }\cdot {\bm\sigma }\right) ,  \label{Ajso}
\end{equation}%
where $q_{j}$ is the momentum component determined by the spin-orbit
coupling strength (with a typical value of the order of $10^{4}$ cm$^{-1}$), 
${\mathbf{h}}^{\left[ j\right] }$ is the direction of the
corresponding spin-orbit coupling field, and ${\bm\sigma }$ is the Pauli
matrix vector. We consider $j-$independent ${\mathbf{h\equiv h}}^{\left[
j\right] }$ corresponding to the single-particle spin-orbit coupling Hamiltonian
in the form
\begin{equation}
H_{\rm so}=\frac{i}{2m}q_{j}{\partial_{j}}\left({\mathbf{h}}\cdot{\bm\sigma}\right),  
\label{Hso}
\end{equation}
such that the spin precession angle caused by the spin-orbit
coupling depends only on the particle displacement along the $\mathbf{q}-$%
direction and the coupling can be gauged out by the corresponding
coordinate-dependent spin rotation around the ${\mathbf h}-$axis. This circumstance 
allows for a mapping of the spin dynamics onto the spin density evolution in
the real space in the absence of the spin-orbit coupling, making it directly
related to the generalized diffusion picture. Such a realization of spin-orbit
coupling has recently been produced for cold three-dimensional 
Fermi gases of $^{40}$K [\onlinecite{Wang}] and $^{6}$Li [\onlinecite{Cheuk}]
and corresponds to formation of spin helix states \cite{Bernevig06,Koralek09,Schliemann16}.
For cold Fermi gases $q$ and the Fermi momentum $p_{F}$ are of the same order 
of magnitude since they both are determined by $2\pi/\lambda,$ where $\lambda$ 
is the wavelength of an optical photon  \cite{Wang,Cheuk,Liu09}.

There are several ways to produce a nonequilibrium spin polarization and track the subsequent 
spin dynamics. One option is to excite a spin density in a certain region of the 
momentum space, as it is done in cold fermions \cite{Wang,Cheuk,Huang16} and in solids \cite{Korn}
and then to study its evolution. Another option is to produce a nonequilibrium 
spin polarization by an external, a uniform or a weakly coordinate-dependent, Zeeman-like
field, let the system reach the equilibrium in the applied field,  
and then to switch this field off. We follow the second approach and consider 
spin polarization initially produced by a weak Zeeman $\delta\mathcal{A}^{0}$
-term, switched off at $t=0.$ Its subsequent evolution is determined by the spin
diffusion kernel in the momentum space in the absence of spin-orbit
coupling, however, taken at the momentum $\mathbf{q}$ in Eq. (\ref{Ajso}) 
\cite{Tokatly10a,Tokatly10b}. With the spin-orbit coupling in Eq. (\ref{Ajso}), the spin projection at the 
${\mathbf{h}}-$axis is conserved, and we will consider evolution of the
orthogonal to ${\mathbf{h}}$ spin component, which we denote as $S(t).$  
The time dependence of the spin can be presented in the
form: 
\begin{equation}
\frac{S(t)}{S(0)}=\frac{\delta s_{\mathbf{q}}(t)}{\delta s_{\mathbf{q}}(0)},
\label{St_S0}
\end{equation}%
where $\delta s_{\mathbf{q}}(t)$ is the corresponding component of the spin
density evolving as in the absence of the spin-orbit coupling. In other
words, the coordinate-spin mapping can be expressed in terms of the exact
frequency and momentum-dependent diagonal spin susceptibility $\chi _{\sigma
\sigma }(\mathbf{q},\omega )$ as 
\begin{equation}
\frac{S(t)}{S(0)}=\int_{-\infty }^{\infty }\frac{d\omega }{2\pi }\left[ 
\frac{\chi _{\sigma \sigma }(\mathbf{q},\omega )}{\chi _{\sigma \sigma }(%
\mathbf{q},0)}-1\right] \frac{e^{-i\omega t}}{i\omega }.  \label{S-general}
\end{equation}%
This approach is valid both for fermionic and bosonic systems in the absence
of a steady spin-related magnetization. Here we apply this mapping in form
of Eqs. (\ref{St_S0}) and (\ref{S-general}) to cold gases in the SU(2)$%
\otimes $U(1)-symmetric fields and study the qualitative features brought
about by the orbital motion in the U(1) pseudomagnetic field.

This paper is organized as follows. In section 2 we introduce for
three-dimensional systems the collective coordinate variables related to the
motion of a particle with the momentum corresponding to the spin-orbit
coupling in a synthetic magnetic field. This allows us to formulate the
corresponding linear response theory to calculate $\delta s_{\mathbf{q}}(t)$
and to study the spin dynamics. This approach is applicable at any strength of
the spin-orbit coupling. In section 3  it will be applied to the spin
dynamics, both in the reversible ballistic and irreversible
collision-dominated regimes. The results will be summarized in Conclusions.
In the Appendices we show how the same results in the collisionless limit can be obtained by considering
the real single-particle trajectories and briefly provide the results for
two-dimensional systems.

\section{Density matrix and auxiliary trajectories}

\begin{figure}[t]
\includegraphics[width=70mm]{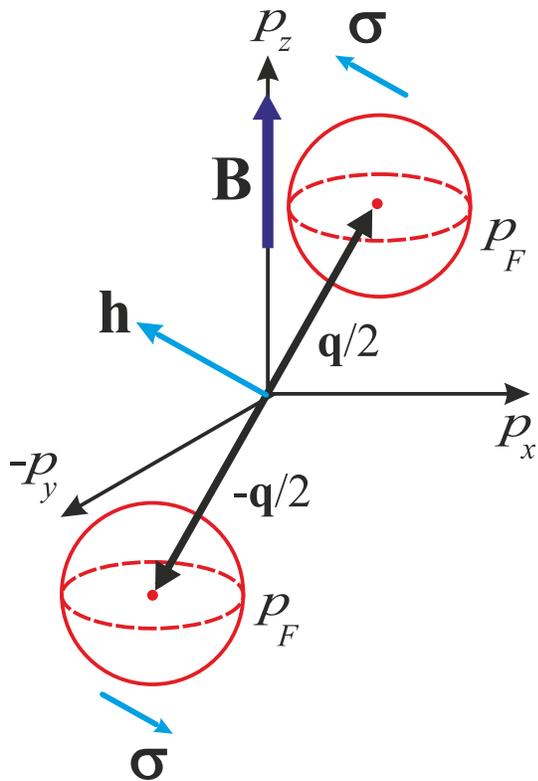}
\caption{Magnetic field $\mathbf{B}$, the spin-orbit coupling momentum $\mathbf{q}$, and the corresponding 
spin-split Fermi surfaces with the Fermi momentum $p_{F}.$ Vector $\mathbf{h}$ is the direction of the 
spin-orbit coupling field in Eq.(\ref{Hso})
and ${\bm\sigma}$ shows the equilibrium magnetization of the components of the spin-split Fermi gas.
This picture corresponds to ultrastrong coupling with $q>p_{F}.$}
\label{geometry}
\end{figure}

To introduce the approach based on the auxiliary trajectories determined by the spin-orbit coupling, we consider
an atomic gas in a uniform synthetic "magnetic" field parallel to the $z-$axis and characterized 
by ${\bm \omega}_{c}=\mathbf{z}\omega_{c},$ where $\omega_{c}=B/m$ is the cyclotron frequency or the
macroscopic rotation frequency \cite{Cooper}, 
as shown in Fig. (\ref{geometry}). {In the following we adopt the Landau gauge to describe the U(1) magnetic field. Our aim is to calculate 
the spin response function $\chi_{\sigma\sigma}({\bf q},\omega)$ entering Eq.~(\ref{S-general})
by solving the equation of motion for the one-particle spin-density matrix $\rho({\bf r}_1,{\bf r}_2,t)$. 
It is convenient to introduce the U(1) gauge invariant density matrix as follows} 
\begin{equation}
f\left( \mathbf{r},\mathbf{R}\right) =\exp \left( i\frac{yx}{l_{B}^{2}}%
\right) \rho \left( \mathbf{r},\mathbf{R}\right) ,
\end{equation}%
where {${\bf r}={\bf r}_1-{\bf r}_2$ and ${\bf R}=({\bf r}_1+{\bf r}_2)/2$ are the relative 
and the center-of-mass coordinates}, $\rho \left( \mathbf{r},\mathbf{R}\right) $ is the density matrix in
the Landau gauge, and $l_{B}$ is the corresponding magnetic length. {The spin density is 
related to the diagonal element of the density matrix ${\bf s}({\bf R})= {\rm Tr}[{\bm\sigma}f\left( 0,\mathbf{R}\right)]$}.

{We consider first a collisionless, purely ballistic dynamics of noninteracting gas. 
This regime of dynamics is realized when the spin precession rate $\langle v\rangle q$ and/or the cyclotron frequency are larger 
than the collisional relaxation rate $1/\tau$ (here $\langle v\rangle$ is the characteristic velocity of particles). The} equation
of motion {for the density matrix} in the presence of external coordinate- and time-dependent
Zeeman-like perturbation $\delta\mathcal{A}^{0}\left( \mathbf{R},t\right),$
corresponding to the field orthogonal to the $\mathbf{h-}$axis, becomes:
\begin{eqnarray}
&&\left[i\partial _{t}-\left(\frac{\widehat{P}_{i}}{m}+\omega_{c}\varepsilon
_{zij}r_{i}\right) \widehat{p}_{j}\right] f\left( \mathbf{r},\mathbf{R}%
\right) = \nonumber\\
&&\quad\left[\delta\mathcal{A}^{0}\left( \mathbf{R+}\frac{\mathbf{r}}{2}%
,t\right) -\delta\mathcal{A}^{0}\left( \mathbf{R-}\frac{\mathbf{r}}{2},t\right)
\right]f\left( \mathbf{r},\mathbf{R}\right) ,
\end{eqnarray}%
where $\widehat{P}_{i}=-i\partial /\partial R_{i},$ $\widehat{p}%
_{j}=-i\partial /\partial r_{j},$ and $\varepsilon _{zij}$ is the
corresponding component of the Levi-Civita fully antisymmetric tensor.

{In the linear response regime the density matrix is weakly perturbed from from its 
equilibrium form $f_{0}\left( \mathbf{r}\right)$}:
\begin{equation}
f\left( \mathbf{r},\mathbf{R}\right) =f_{0}\left( \mathbf{r}\right) +\delta
f\left( \mathbf{r},\mathbf{R}\right).
\end{equation}%
{The linearized version of the equation of motion reads:}
\begin{eqnarray}
&&\left[ i\partial _{t}-\left(\frac{\widehat{P}_{i}}{m}+\omega_{c}\varepsilon
_{zij}r_{i}\right) \widehat{p}_{j}\right] \delta f\left( \mathbf{r},\mathbf{%
R}\right) = \nonumber \\
&&\quad\left[\delta\mathcal{A}^{0}\left( \mathbf{R+}\frac{\mathbf{r}}{2}%
,t\right) -\delta\mathcal{A}^{0}\left( \mathbf{R-}\frac{\mathbf{r}}{2},t\right)
\right] f_{0}\left( \mathbf{r}\right).  \label{general}
\end{eqnarray}%

Since we are interested in the ${\bf q}$-dependent response it is sufficient to consider a plane-wave perturbation 
\begin{equation}
\delta\mathcal{A}^{0}\left( \mathbf{r},t\right) =\delta\mathcal{A}^{0}_{\mathbf{q}}\left(
t\right) e^{i\mathbf{q}\cdot\mathbf{r}}.
\end{equation}%
{This perturbation induces the response of} the form: $\delta f\left( \mathbf{r},\mathbf{R},t\right)
=\delta f_{_{\mathbf{q}}}\left( \mathbf{r},t\right) e^{i\mathbf{q}\cdot\mathbf{r}}.$
By substituting $\delta\mathcal{A}^{0}\left( \mathbf{r},t\right) =
\mathcal{A}^{0}_{\mathbf{q}}\left( t\right) e^{i\mathbf{q}\cdot\mathbf{r}}$ and $\delta f\left( \mathbf{r},%
\mathbf{R},t\right) =\delta f_{_{\mathbf{q}}}\left( \mathbf{r},t\right) e^{i%
\mathbf{q}\cdot\mathbf{r}}$ into Eq.~(\ref{general}) we obtain {the following final equation of motion}%
\begin{eqnarray}
&&\left[ i\partial _{t}-\left( \frac{q_{i}}{m}+\omega _{c}\varepsilon
_{zij}r_{i}\right) \frac{\partial }{\partial r_{j}}\right] 
\delta f_{_{\mathbf{q}}}\left( \mathbf{r},t\right) = \nonumber \\
&& \quad 2\delta\mathcal{A}^{0}_{\mathbf{q}}\left(
t\right) 
\sin\left[\frac{\mathbf{q}\cdot\mathbf{r}}{2}\right]
f_{0}\left(\mathbf{r}\right). 
\label{Boltzmann}
\end{eqnarray}%
The left-hand-side of
this equation is the Boltzmann operator, corresponding to the helix-like motion of
particles having velocities $\mathbf{q}/m$ in a magnetic field. The
right-hand side of this equation plays the role of the source of these
particles \cite{Kirzhnitz}. To understand the role of the $q_{i}-$ parameters in the 
helical motion, we introduce  the anomalous spin-dependent term in the velocity operator 
\begin{equation}
\mathbf{v}_{\mathrm{\rm so}}\equiv i\left[H_{\mathrm{\rm so}},\mathbf{r}\right]=-\frac{\mathbf{q}}{2m}
\left({\mathbf{h}}\cdot{\bm\sigma}\right),
\end{equation}
and find that $\mathbf{q}/m$ is the difference of the eigenvalues of $\mathbf{v}_{\rm so}$ corresponding to 
${\bm\sigma}$ antiparallel and parallel to $\mathbf{h}.$

To {solve} Eq. (\ref{Boltzmann}) we introduce auxiliary
trajectories ${\bm\xi}(\mathbf{r},t),$ satisfying the following
equation of motion,  
\begin{equation}\label{xi-eom}
\overset{.}{\mathbf{\xi }_{j}}(\mathbf{r},t)=\frac{q_{i}}{m}+\omega
_{c}\varepsilon _{zij}{\mathbf{\xi }_{i}}(\mathbf{r},t),
\end{equation}%
{with the} initial condition ${\mathbf{\xi }_{j}}(\mathbf{r},t)=r_{j}.$ {The trajectory ${\bm\xi}(\mathbf{r},t),$ 
corresponds to a particle moving from the point $\mathbf{r}$ with the initial momentum $\mathbf{q}$ in 
the presence of a given magnetic field.} 
{By a direct substitution one easily checks that} the solution {to Eq.~(\ref{Boltzmann})} can be expressed as:%
\begin{eqnarray}
&&\delta f_{_{\mathbf{q}}}\left( \mathbf{r},t\right) = 2\times\\
&& \int_{-\infty
}^{0}dt_{1}\delta\mathcal{A}^{0}_{\mathbf{q}}\left( t_{1}+t\right) 
\sin\left[\frac{\mathbf{q}\cdot {\bm\xi }(\mathbf{r},t_{1})}{2}\right]
f_{0}\left(\bm\xi(\mathbf{r},t_{1})\right). 
\nonumber
\end{eqnarray}%
Taking into account that the spin density response corresponds to the density matrix at $\mathbf{r}=0$, {i.~e. 
$\delta s_{_{\mathbf{q}}}\left( t\right)=\delta f_{_{\mathbf{q}}}\left( 0,t\right)$,} we
obtain 
\begin{eqnarray} \label{spin-1}
&&\delta s_{_{\mathbf{q}}}\left( t\right) = 2\times \\
&& \int_{-\infty }^{0}dt_{1}\mathcal{A}^{0}_{\mathbf{q}}\left( t_{1}+t\right) 
\sin \left[\frac{\mathbf{q}\cdot{\bm\xi}(t_{1})}{2}\right] 
f_{0}\left({{\bm\xi}}(t_{1})\right), \nonumber
\end{eqnarray}%
{where we introduced the} notation ${{\bm\xi}}(t_{1})\equiv{{\bm\xi}}({0},t_{1}).$ 

Without loss of generality the
spin-orbit coupling momentum can be taken as ${\mathbf{q}}=\left( {q}_{\perp
},0,q_{\parallel }\right)$, where $q_{\parallel}=q\cos\theta$ and $q_{\perp}=q\sin\theta$ with $\theta$ being 
the angle between the direction of ${\bf q}$ and the magnetic field. In this parametrization 
the auxiliary helix trajectories take the following explicit form: 
\begin{eqnarray}
\xi _{x}(t) &=&\frac{q_{\perp }}{m\omega _{c}}\sin \left( \omega
_{c}t\right) ,\quad \xi _{y}(t)=\frac{q_{\perp }}{m\omega _{c}}\left[
1-\cos \left( \omega _{c}t\right) \right]   \nonumber \\
\xi _{z}(t) &=&\frac{q_{\parallel }}{m}t,  \nonumber \\
\xi (t) &=&\sqrt{4\left( R_{q}^{\perp }\right) ^{2}\sin ^{2}\frac{\omega
_{c}t}{2}+\left( \frac{q_{\parallel }}{m}t\right) ^{2}},
\label{xi}
\end{eqnarray}%
where $R_{q}^{\perp }=q_{\perp }/m\omega _{c}$ and%
\begin{equation}
\mathbf{q}\cdot {\bm\xi}(t) =\frac{q_{\perp }^{2}}{m\omega _{c}}%
\sin \left( \omega _{c}t\right) +\frac{q_{\parallel }^{2}}{m}t.
\end{equation}

{Up to this point we specify neither statistics of the particles, nor the temperature. 
All this information is encoded in the} equilibrium density matrix $f_{0}\left( \mathbf{r}\right)$. 
{Below we consider the experimentally relevant quasiclassical limit with respect to the U(1) 
magnetic field, that is} $\omega _{c}\ll \left\langle E\right\rangle$, where $%
\left\langle E\right\rangle$ is the characteristic particle
energy. {In this limit $f_{0}\left( \mathbf{r}\right)$} reduces to the Fourier transform of the equilibrium distribution
function $f(p)$: 
\begin{equation}
f(p)=\frac{1}{\exp \left[ \left( p^{2}/2m-\mu \right) /T\right] \pm 1},
\label{f_p}
\end{equation}
where $\pm$ corresponds to the Fermi/Bose statistics of the particles.

\section{Collisionless motion and spin relaxation}

\subsection{General solution}

The special feature of a three-dimensional gas is that for a nonzero $%
q_{\parallel }$ {the trajectories ${\bm\xi}({\bf r},t)$ are open, and} a particle can move infinitely far from the initial
position, making the behavior qualitatively different from that expected in a two-dimensional gas. 

For the subsequent discussion of spin dynamics it is convenient to represent the Fourier transform of the distribution 
function {in the following form}%
\begin{equation} \label{f_0-N}
f_{0}\left( \mathbf{r}\right) =\int \frac{d^{3}p}{\left( 2\pi \right) ^{3}}%
e^{i\mathbf{p}\cdot\mathbf{r}}f(p)=-\frac{1}{\pi r}\frac{d}{dr}\mathcal{N}(r),
\end{equation}%
where 
\begin{equation} \label{N-def}
\mathcal{N}(r)\equiv \int_{0}^{\infty }\cos (pr)f(p)\frac{dp}{2\pi }
\end{equation}%

The function $\mathcal{N}({\xi }(t))$ will play an important role in the following analysis. 
First of all we notice that $\lim_{t\rightarrow \infty }\mathcal{N}({\xi }(t))_{q_{\parallel \neq 0}}=0$, 
{which follows from the openness of the particle trajectories.}
We {also} show the functions $\mathcal{N}(\xi(t))$ for two {limiting cases} of interest. For a
degenerate Fermi gas {with $f(p)=\theta(p_F -p)$ we find:}
\begin{equation} \label{N-Fermi}
\mathcal{N}({\xi }(t))=\int \frac{dp}{2\pi }\theta \left( p_{F}-p\right)
\cos (p{\xi }(t))=\frac{1}{2\pi }\frac{\sin \left( p_{F}\xi (t)\right) }{\xi
(t)}.
\end{equation}%
For highly degenerate Bose gases slightly above to the Bose-Einstein condensation
transition, where the chemical potential $-\mu $ is much less than the
temperature $T$, the low-momentum part of the distribution function 
{reduces to the form} $f(p)=2p_{T}^{2}/\left(
p^{2}+p_{\mu }^{2}\right)$, where $p_{\mu }^{2}\equiv 2m\left| \mu
\right| \ll p_{T}^{2}$ and $p_{T}^{2}\equiv mT$. {In this limit we obtain:}  
\begin{equation}
\mathcal{N}({\xi}(t))=\int \frac{dp}{2\pi}\frac{2p_{T}^{2}}{p^{2}+p_{\mu
}^{2}}\cos (p{\xi}(t))=\frac{p_{T}^{2}}{2p_{\mu}}\exp(-p_{\mu}{\xi}(t)).
\label{NBose}
\end{equation}%
Note that Eq. (\ref{NBose}), based on the small-momentum low-$T$ 
form of the Bose distribution, is exact only at sufficiently large displacements 
such that $p_{T}{\xi}(t)\gg 1$ [\onlinecite{small_t_bosons}].  

{Now we proceed further with the transformation of the density matrix $f_{0}\left(\bm\xi(t)\right)$ entering Eq.~(\ref{spin-1})}. 
{Using the trajectory equation (\ref{xi-eom}) we find the following} identity:%
\begin{equation}
\frac{d}{dt}{\xi }(t)=\frac{\overset{.}{{\bm\xi }}(t)\cdot {\bm\xi}(t)}{{\xi }(t)}=\frac{\mathbf{q}\cdot {\bm\xi }(t)}{m{\xi }(t)}.
\end{equation}%
{Making use of this identity} we obtain 
\begin{equation*}
\frac{d\mathcal{N}({\xi }(t))}{d{\xi }(t)}=\frac{m{\xi }(t)}{\mathbf{q%
}\cdot {\bm\xi }(t)}\frac{d\mathcal{N}({\xi }(t))}{dt}.
\end{equation*}%
{Finally, using the representation of Eq.~(\ref{f_0-N}) we find for the equilibrium density matrix 
evaluated at the particle trajectory:} \begin{equation}
f_{0}\left( {\bm\xi }(t)\right) =-\frac{1}{\pi {\xi }(t)}\frac{d\mathcal{N}(\xi
(t))}{d{\xi }(t)}=-\frac{m}{\pi }\frac{1}{\mathbf{q}\cdot{\bm\xi}%
(t)}\frac{d\mathcal{N}({\xi }(t))}{dt}.
\end{equation}%
As a result, the spin density response to the perturbation $\mathcal{A}^{0}_{\mathbf{q}}\left( t_{1}\right) $ takes the form%
\begin{eqnarray} \label{spin-2}
&&\delta s_{_{\mathbf{q}}}\left( t\right) = \\
&&\frac{2m}{\pi}
\int_{-\infty}^{0}dt_{1}\delta\mathcal{A}^{0}_{\mathbf{q}}\left( t_{1}+t\right) 
\frac{\sin \left[\displaystyle
{\frac{\mathbf{q}\cdot {\bm\xi }(t_{1})}{2}}
\right]}{\mathbf{q}\cdot {\bm\xi }(t_{1})}
\frac{d\mathcal{N}({\xi }(t_{1}))}{dt_{1}}. \nonumber
\end{eqnarray}%
Below we apply this equation to study the reversible collisionless spin evolution
and spin relaxation in the collision-dominated regime.

\subsection{Collisionless spin dynamics as switch-off response}

\begin{figure}[t]
\includegraphics*[width=70mm]{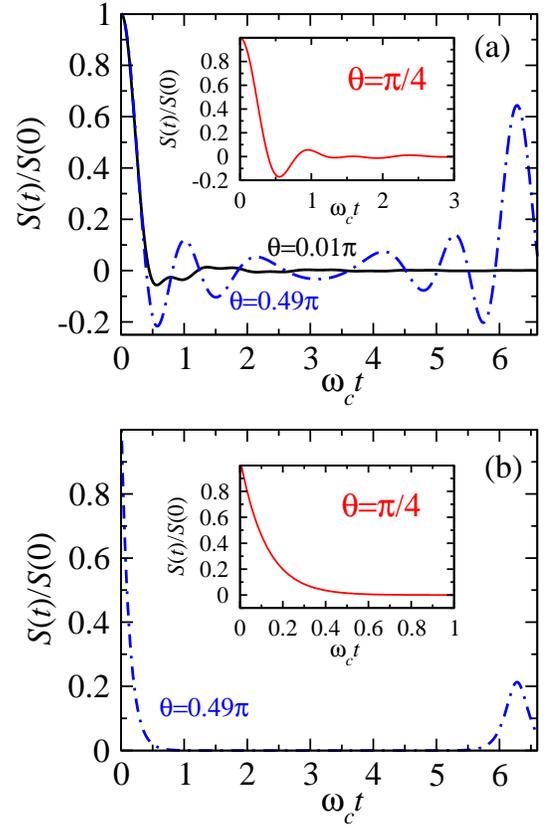}
\caption{Time-dependence of the total spin calculated with exact Eq.(\ref{exact}) for different values
of $\theta$ as marked near the plots.  (a) For fermions with $p_{F}\equiv 1$, (b) for bosons, 
with $p_{\mu}\equiv 1$. Here we use $q=0.5,$ $\omega_{c}=1/16,$ and $m\equiv 1$. 
An approximation with Eq. (\ref{finiteq}) (not shown here) describes all evolutions very accurately. 
The insets show the evolution for $\theta=\pi/4.$}
\label{spinevolution}
\end{figure}

In general to find the spin evolution from Eq.~(\ref{S-general}) one needs the spin response 
function $\chi_{\sigma\sigma}({\bf q},\omega)$. The latter requires calculation of the 
response to a time periodic perturbation $\sim e^{-i\omega t}$. However 
the reversible collisionless spin dynamics can be extracted directly from Eq.~(\ref{spin-2}) 
by considering the switch-off process with the driving field of the following form 
\begin{equation}
\delta\mathcal{A}^{0}_{\mathbf{q}}\left( t\right) =\exp \left( \delta t\right)\theta(-t),
\end{equation}%
where the limit $\delta \rightarrow +0$ is assumed. In other words, the Zeeman field 
is switched on adiabatically to polarize the system (in the 
equilibrium at $t<0$), and then switched off at $t=0$ 
and follow the dynamics for $t>0$. The corresponding
spin density response acquires the form:%
\begin{equation}
\delta s_{_{\mathbf{q}}}\left( t\right) =\frac{2m}{\pi }\int_{t}^{\infty}dt_{1}
\frac{\sin \left[\displaystyle
{\frac{\mathbf{q}\cdot {\bm\xi}(t_{1})}{2}}\right]}{\mathbf{q}\cdot{\bm\xi}(t_{1})}
\frac{d\mathcal{N}(%
{\xi }(t_{1}))}{dt_{1}}. \label{exact}
\end{equation}%

{The characteristic scale of the function $\mathcal{N}(p)$ 
is determined by the mean velocity $\langle v\rangle$. This allows 
for the further simplification of Eq.~(\ref{exact}) in the limit $\langle v\rangle\gg q/m$.} 
Integrating by parts and separating the fast motion we obtain {the approximate expression}, 
formally valid for small {spin-orbit} momentum $q$:%
\begin{equation}
\frac{S(t)}{S(0)}=
2\frac{\sin \left[\displaystyle
{\frac{\mathbf{q}\cdot{\bm\xi}(t)}{2}}
\right]}{\mathbf{q}\cdot{\bm\xi}(t)}
\frac{\mathcal{N}({\xi }(t))}{\mathcal{N}(0)}.  \label{finiteq}
\end{equation}

The results of calculation with the exact Eq.(\ref{exact}) are presented in
Fig.~\ref{spinevolution} for Fermi and Bose statistics. For $\theta$ close to $\pi /2$ one expects revivals of the spin density after a
cyclotron period: $S(2\pi\,n/\omega _{c})=S(0).$ These revivals, however, are
strongly suppressed at smaller angles by the motion along the magnetic field, as
can be seen in Fig.~\ref{spinevolution}.

It is interesting to note that although approximation of Eq.~(\ref{finiteq}) is formally 
valid in the small-$q$ limit, it describes the spin evolution 
highly accurately even for $q\sim m\langle v\rangle$. Therefore the main features of the reversible dynamics can be understood 
from the analytic formula Eq.~(\ref{finiteq}). Let us consider first the degenerate Fermi gas with $\mathcal{N}(\xi)$ given by Eq.~(\ref{N-Fermi}) 
and substitute the trajectory of Eq.~(\ref{xi}). Then Eq. (\ref{finiteq}) acquires the following explicit form 
\begin{eqnarray}
\frac{S(t)}{S(0)} &=&2\frac{\sin \left( p_{F}\sqrt{4\left( R_{q}^{\perp
}\right) ^{2}\sin ^{2}\displaystyle{\frac{\omega _{c}t}{2}}+\left(\displaystyle{\frac{q_{\parallel }}{m}}
t\right) ^{2}}\right) }{p_{F}\sqrt{4\left( R_{q}^{\perp }\right) ^{2}
\sin^{2}\displaystyle{\frac{\omega _{c}t}{2}}+\left(\displaystyle{\frac{q_{\parallel}}{m}}t\right) ^{2}}}
\nonumber \\
&&\times \frac{\sin \left[ \displaystyle{\frac{1}{2}}
\left(\displaystyle{\frac{q_{\perp }^{2}}{m\omega_{c}}}\sin 
\left(\omega _{c}t\right)+\displaystyle{\frac{q_{\parallel }^{2}}{m}}t\right) %
\right] }{\displaystyle{\frac{q_{\perp }^{2}}{m\omega _{c}}}\sin \left( \omega _{c}t\right)
+\displaystyle{\frac{q_{\parallel }^{2}}{m}}t}.
\label{Fermi} 
\end{eqnarray}%
{This expression shows that} the initial stage is well-described in the limit of a straight motion
with $\xi (t)=qt/m,$ producing a universal evolution which is $\omega_{c}$ and $%
\theta -$independent. Taking into account the short-time expansion $qt/m-\xi
(t)=qt/m\times \left( \omega _{c}t\right) ^{2}\sin ^{2}\theta /24,$ we
obtain that the cyclotron motion modifies this universal behavior at times $%
q\left( p_{F}/m\right) t\sin ^{2}\theta \left( \omega _{c}t\right)
^{2}/24\sim 1,$ that is $t\sim 3/\omega _{c}\left( \sin ^{2}\theta
p_{F}q/m\omega_{c}\right) ^{1/3}.$ Note that if the angle $\theta $ is
sufficiently small, that is $\theta \ll \left(m\omega_{c}/p_{F}q\right)^{1/2},$ the
cyclotron motion has only a weak influence on the spin dynamics since the
effective displacement along the field direction is always sufficiently
larger that the cyclotron radius. Another timescale of the problem is
determined by the second, $p_{F}-$independent factor in Eq.(\ref{Fermi}) At $%
\theta =0,$ it leads to additional decay factor in the spin density as $\sin
\left( q_{\parallel }^{2}t/2m\right) /\left( q_{\parallel }^{2}t/m\right) ,$
while at $\theta =\pi /2$ it leads to a modulation of the $\sin
\left( p_{F}qt\right) /p_{F}qt$ decay. The revivals show a moderate
suppression by the along-the field motion as $\sin (p_{F}nq_{\parallel
}t_{c})/p_{F}nq_{\parallel }t_{c}.$

The same arguments can be applied to the low-temperature Bose gas \cite{nondegenerate} {with $\mathcal{N}(\xi)$ of Eq.~(\ref{NBose})}. 
{In this case Eq.~(\ref{finiteq}) reduces to the form}  
\begin{eqnarray}
\frac{S(t)}{S(0)} &=&2\exp \left( -p_{\mu }\sqrt{4\left( R_{q}^{\perp
}\right) ^{2}\sin ^{2}\frac{\omega _{c}t}{2}+\left( \frac{q_{\parallel }}{m}%
t\right) ^{2}}\right) \nonumber\\
&&\times \frac{\sin \left[ \displaystyle{\frac{1}{2}}
\left(\displaystyle{\frac{q_{\perp }^{2}}{m\omega_{c}}}\sin \left( \omega _{c}t\right)+\displaystyle{\frac{q_{\parallel }^{2}}{m}}t\right) %
\right] }{\displaystyle{\frac{q_{\perp }^{2}}{m\omega _{c}}}\sin \left( \omega _{c}t\right)
+\displaystyle{\frac{q_{\parallel }^{2}}{m}}t}.
 \label{Bose} 
\end{eqnarray}%
Because of a broad distribution of the momenta, the revivals at full
cyclotron periods here are suppressed much stronger than for the Fermi gas, that
is exponentially as $\exp (-np_{\mu }q_{\parallel }t_{c})$, which is clearly visible in Fig.~\ref{spinevolution}b.

\subsection{Susceptibility, diffusion, and spin relaxation in
collision-dominated limit}

Equation (\ref{exact}) describes collisionless reversible spin dynamics 
which originate from the inhomogeneous spin precession of particles moving 
ballistically along helicoidal trajectories in the magnetic field. 
Now we consider the opposite limit of collision-dominated diffusive 
dynamics corresponding to the situation when the spin precession rate is 
smaller than the rate of collisional relaxation $q\langle v\rangle\ll 1/\tau$. 
To take into account the effect of collisions we apply our general formula 
Eq.~(\ref{S-general}). For the response function $\chi_{\sigma\sigma}({\bf q},\omega)$ 
we use the density-conserving relaxation time approximation that has been introduced 
by Mermin  \cite{Mermin} and adapted recently \cite{Tokatly13} to the problems
of spin dynamics. In this approximation the spin response function is calculated as 
\begin{equation}
\chi _{\sigma \sigma }\left( \mathbf{q,}\omega \right) =\frac{\chi
_{0}\left( \mathbf{q},\omega +i/\tau \right) }
{1-\displaystyle{\frac{1}{1-i\omega\tau}}
\left[1-\displaystyle{\frac{\chi _{0}\left(\mathbf{q},\omega +i/\tau \right)}{\chi _{0}\left(\mathbf{q},0\right)}}\right]},  
\label{Mermin_chi}
\end{equation}
where $\chi_{0}({\bf q},\omega)$ is a bare susceptibility of the system in the absence of collisions. 
The latter quantity is obtained straightforwardly from Eq.~(\ref{spin-2}) by considering the 
response to a time-periodic Zeeman
field of the form $\delta\mathcal{A}^{0}_{\mathbf{q}}\left(t\right)=\delta\mathcal{A}^{0}_{\mathbf{q}}e^{-i\omega t}$: 
\begin{equation} 
\chi _{0}\left( \mathbf{q,}\omega \right) =\frac{2m}{\pi }\int_{0}^{\infty
}dt e^{i\omega t}
\frac{\sin \left[\displaystyle
{\frac{\mathbf{q}\cdot {\bm\xi}(t)}{2}}
\right]}{\mathbf{q}\cdot {\bm\xi }(t)}
\frac{d\mathcal{N}({\xi }(t))}{dt}.
\label{chi_0-general}
\end{equation}%
Equations (\ref{S-general}), (\ref{Mermin_chi}), and (\ref{chi_0-general}) describe the spin 
dynamics for any $q$, $\omega_c$, and $\tau$, provided the magnetic field can be treated semiclassically. 

In the collision-dominated regime only small frequencies $\omega\ll 1/\tau$ 
contribute to the integral in Eq.~(\ref{S-general}). In this limit the integrand in Eq.~(\ref{S-general}) 
acquires a simple pole on the imaginary axis
\begin{eqnarray}
 \label{diffusion}
&&\frac{1}{\omega }\left[ 
\frac{\chi _{\sigma \sigma }(\mathbf{q},\omega )}{\chi _{\sigma \sigma }(%
\mathbf{q},0)}-1\right]  \approx
\frac{1}{\omega + i\Omega({\bf q})}, \\
&&\Omega({\bf q}) = \frac{1}{\tau}\frac{\chi_0({\bf q},i/\tau)}{\chi_0({\bf q},0)},
\label{Omega}
\end{eqnarray}
which is a clear signature of the diffusive dynamics \cite{Forster,TYu}. To find the position of the diffusive 
pole we notice that collision-dominated regime occurs only in the small-$q$ limit. 
Therefore a simplified form of the bare susceptibility Eq.~(\ref{chi_0-general}) can be used as
\begin{eqnarray} \nonumber
&&\chi _{0}\left( \mathbf{q,}\omega \right) = \frac{m}{\pi }\int_{0}^{\infty}dt 
e^{i\omega t}\frac{d\mathcal{N}({\xi }(t))}{dt} \\
&=& -\frac{m}{\pi }\mathcal{N}(0) -i\omega\frac{m}{\pi }\int_{0}^{\infty}dt e^{i\omega t}\mathcal{N}({\xi }(t))
\label{chi_0-smallq}.
\end{eqnarray}%
Using this expression and the definition of $\mathcal{N}({\xi }(t))$ in Eq.~(\ref{N-def}) we find that $\chi _{0}\left( \mathbf{q},0 \right)$ 
can be expressed in terms of the derivative of the density $n$ of particles with respect to the chemical potential,
\begin{equation}
 \label{chi_0-0}
\chi _{0}\left( \mathbf{q},0 \right)=-\frac{\partial n}{\partial\mu},
\end{equation}
in agreement with the compressibility sum rule \cite{Vignale} corresponding to an adiabatic modification 
of the density by a smooth external potential. By substituting $\omega=i/\tau$ into Eq.~(\ref{chi_0-smallq}) 
and expanding $\mathcal{N}(\xi)$ to the second order in $\xi$ we compute $\chi _{0}\left(\mathbf{q},i/\tau\right)$ 
in the diffusive limit
\begin{eqnarray}\nonumber
\chi _{0}\left( \mathbf{q,}i/\tau \right) &=&-\frac{mn}{2\tau}
\int_{0}^{\infty }\exp \left(-t/\tau \right){\xi }^{2}(t)dt \\
&=& -\frac{\tau ^{2}n}{m}\left(
q_{\parallel }^{2}+\frac{q_{\perp }^{2}}{1+\omega _{c}^{2}\tau ^{2}}\right).
\label{chi_0-tau}
\end{eqnarray}%
In the second line of this equation we used the explicit form of $\xi(t)$ for the helix trajectory, Eq.~(\ref{xi}).
The anisotropic $\mathbf{q}$-dependence reflects the fact that at long times the circular motion is
restricted as $\xi _{\perp }^{2}\leq 4\left(R_{q}^{\perp}\right)^{2}\sim \left( q_{\perp
}/m\omega _{c}\right) ^{2},$ and, therefore, cannot give a large
contribution into the spin evolution, while the parallel motion with $%
\xi _{\parallel }^{2}=\left( q_{\parallel }/m\right) ^{2}t^{2}$ is unrestricted and influences
this evolution much stronger.

Equations (\ref{Omega}), (\ref{chi_0-0}), and (\ref{chi_0-tau}) determine the position of the diffusive pole 
in the response function
\begin{equation}
 \Omega({\bf q}) = D\left(
q_{\parallel }^{2}+\frac{q_{\perp }^{2}}{1+\omega _{c}^{2}\tau ^{2}}\right),
\end{equation}
where $D=\tau{n}\left({\partial \mu}/{\partial n}\right)/{m}$ is the diffusion coefficient. 
Finally, Eqs.~(\ref{S-general}) and (\ref{diffusion}) 
yield the modified Dyakonov-Perel spin relaxation \cite{Dyakonov73,Ivchenko,Burkov} 
dependent on $\mathbf{q}$-direction as follows,
\begin{equation}
\frac{S(t)}{S(0)}=\exp \left[ -D\left( q_{\parallel }^{2}+
\frac{q_{\perp}^{2}}{1+\omega _{c}^{2}\tau ^{2}}\right) t\right].
\label{StS0_DP}
\end{equation}%
Importantly, the statistics of the particles and the temperature enter this equation only through 
the diffusion coefficient. 
Qualitatively in the collision-dominated regime the behavior of the spin is universal \cite{D_bosons}. 
For $q_{\perp }=0$, the magnetic field does not influence the particle
trajectory, and we return to the conventional Dyakonov-Perel relaxation
in zero magnetic field. In the case of a strong magnetic field
and $q_{\parallel}=0,$ the relaxation rate can be understood as follows.
Particles move over circular trajectories of a small radius $R_{c}=v/\omega_{c}$ 
with $R_{c}\ll 1/q.$ The trajectory relatively rarely (after the collision time $\tau \gg
1/\omega_{c}$) experiences considerable changes by displacement of the order of $R_{c}.$  
Therefore, the guiding center of the trajectory diffuses
with characteristic time-dependent displacement $\rho _{\rm gc}^{2}(t)\sim
R_{c}^{2}\times \left( t/\tau \right).$ Taking into account that $%
R_{c}^{2}\sim D/\omega _{c}^{2}\tau ,$ the condition 
$\rho _{\rm gc}(t_{\rm sr})\sim 1/q$ [\onlinecite{Tokatly10a}] yields the spin 
relaxation time $t_{\rm sr}$ in agreement with
the above Eq.~(\ref{StS0_DP}).

\section{Conclusions}

We have studied (pseudo)spin evolution in cold atomic gases, fermionic and bosonic, in synthetic SU(2)$\otimes$U(1) fields.
Our approach, based on the representation of motion in terms of auxiliary trajectories satisfying the kinetic equation,
utilizes the mapping of spin dynamics onto the spin density
evolution in systems without spin-orbit coupling. This mapping is possible for the 
considered here spin-orbit coupling determined by 
a single momentum-independent precession axis.  
With this mapping, valid for any strength of the spin-orbit coupling, we describe reversible 
collisionless evolution as well as irreversible spin relaxation in 
the collision-dominated limit.

The collisionless limit shows the dependence of the reversible spin evolution on the angle between 
the synthetic U(1) field  $\mathbf{B}$ and the single momentum $\mathbf{q}$ characterizing the SU(2)-coupling. 
For the angle between this spin-orbit coupling momentum and the synthetic magnetic field close to $\pi/2$, the 
total spin shows periodic behavior with returns after each cyclotron period. The periodicity 
disappears at smaller angles due to the motion of atoms along the $\mathbf{B}-$field. 
At low temperatures, the spin dynamics of Fermi gases tends to the 
limit determined by the Fermi-momentum, spin-orbit coupling, and 
the strength of the synthetic magnetic field. At a sufficiently strong spin-orbit coupling, this evolution 
can show an oscillating behavior determined by the spin precession. For the Bose-statistics the oscillations are smeared out
by the thermal Bose distribution.  In addition, the spin evolution of bosons vanishes
due to accumulation of the particles in the vicinity of the zero-energy point at temperatures slightly above the 
Bose-Einstein condensation temperature. 

In the collision-dominated limit the spin relaxation described by anisotropic Dyakonov-Perel
formula, depends on the angle between the momentum corresponding to the spin-orbit 
coupling and the synthetic magnetic field. Here the spin relaxation rate is proportional to the density 
diffusion coefficient. At a given momentum relaxation time, the diffusion coefficient for fermions reaches 
a finite zero-temperature limit proportional to the Fermi energy. 
However, for bosons this diffusion coefficient decreases to zero, and the spin relaxation slows
down as the system approaches the Bose-Einstein condensation.

\begin{acknowledgments}

{IVT acknowledges funding by the Spanish MINECO (FIS2013-46159-C3-1-P) and
\textquotedblright Grupos Consolidados UPV/EHU del Gobierno
Vasco\textquotedblright\ (IT-578-13)}. This work of EYS was supported by the
University of Basque Country UPV/EHU under program UFI 11/55, Spanish MEC
(FIS2012-36673-C03-01 and FIS2015-67161-P), and \textquotedblright Grupos Consolidados UPV/EHU
del Gobierno Vasco\textquotedblright\ (IT-472-10).

\end{acknowledgments}

\appendix

\section{Real single-particle trajectories}

Here we will show with an example that the same results for the collisionless dynamics can
be obtained in the weak spin-orbit coupling limit by using the real rather than auxiliary helix trajectories for
particles in magnetic field. This trajectory for a particle with initial velocity $%
\mathbf{v}=\left( \mathbf{v}_{\perp },v_{z}\right) ,$ where $v_{\perp }$ $%
=v\sin\theta,$ $v_{z}$ $=v\cos\theta$ satisfies conditions $%
z=vt\cos \theta $ and $\rho =2R_{c}\sin \theta \sin \left|{\omega_{c}t}/{2}\right|,$ 
where $\rho $ is the distance from the origin. As a
result, we obtain the collisionless limit for the diffusion kernel in the real space as 
\begin{equation}
\mathcal{D}(\rho,z;t)=\delta \left[ z-vt\cos \theta \right] 
\frac{\delta \left[ \rho -2R_{c}\sin \theta 
\sin\left|\displaystyle{\frac{\omega_{c}t}{2}}\right|\right] }{4\pi R_{c}\sin \theta \left| \sin 
\displaystyle{\frac{\omega_{c}t}{2}}\right|},
\end{equation}%
satisfying the normalization condition%
\begin{equation}
\int \mathcal{D}(\rho ,z;t)d^{2}\rho dz=1.
\end{equation}%
According to Ref. [\onlinecite{Tokatly10b}], taking Fourier transform of $\mathcal{D}(\rho ,z;t)$ at $%
\mathbf{q}=(q_{\perp },0,q_{\parallel })$ and averaging over the Fermi surface, that is over the angles $\theta,$ we arrive at:

\begin{equation}
\frac{S(t)}{S(0)}=\frac{\sin \sqrt{\left( q_{\parallel }vt\right)
^{2}+\left( 2q_{\perp }R_{c}\sin\left(\displaystyle{\frac{\omega _{c}t}{2}}\right)\right)^{2}}}
{\sqrt{\left(q_{\parallel }vt\right) ^{2}+\left( 2q_{\perp }R_{c}
\sin\left(\displaystyle{\frac{\omega _{c}t}{2}}\right)\right)^{2}}},
\end{equation}%
same as Eq.(\ref{Fermi}) for $v=p_{F}/m$.

\section{Two-dimensional gases}

The approach developed in this paper is valid for two-dimensional gases
as well. For completeness, we present here the collisionless time dependence of
spin for three main realizations of statistics while in the collision-dominated regime the 
relaxation is described by the modified Dyakonov-Perel formula.

For degenerate Fermi gas the evolution 
\begin{eqnarray}
&&\frac{S(t)}{S(0)}=\frac{2m\omega_{c}}{q^{2}\sin \left(\omega_{c}t\right) }\sin \left[ \frac{%
q^{2}\sin \left(\omega_{c}t\right) }{2m\omega_{c}}\right]\times \nonumber \\
&&\quad J_{0}
\left(2\frac{p_{F}q}{m\omega_{c}}\sin\frac{\omega_{c}t}{2}\right),   
\label{coll3}
\end{eqnarray}%
in the limit of a weak spin-orbit coupling and U(1) field shows a slower oscillating decrease ($\sim 1/\sqrt{t}$) than in the 3D gas
since in 3D the absolute values of the in-plane components of the Fermi momentum are distributed over the Fermi sphere
between $0$ and $p_{F}$ while in 2D all particles have the same $p_{F}$. In the weak coupling limit this expression
coincides with the result of Glazov in Ref. [\onlinecite{Glazov07}]. 

For a highly degenerate Bose gas close to the Bose-Einstein condensation one obtains
\begin{eqnarray}
&&\frac{S(t)}{S(0)}=\frac{2p_{\mu }}{q}\sin \left[ \frac{q^{2}\sin 
\left(\omega_{c}t\right) }{2m\omega_{c}}\right] \frac{1}{\cos \left(\omega_{c}t/2\right) } \times \nonumber \\
&&\quad K_{1}\left[ 2\displaystyle{\frac{qp_{\mu}}{m\omega_{c}}}\sin \frac{\omega_{c}t}{2}\right],
\end{eqnarray}%
with $K_{1}$ being the Bessel function of the second kind.

For nondegenerate gases with $f(p)=\exp (-p^{2}/2p_{T}^{2})$ the evolution is described by%
\begin{eqnarray}
&&\frac{S(t)}{S(0)}=\frac{2m\omega_{c}}{q^{2}\sin \left(\omega_{c}t\right) }
\sin\left[\frac{q^{2}\sin \left(\omega_{c}t\right) }{2m\omega_{c}}\right] \nonumber \times \\
&&\quad \exp \left[-\frac{q^{2}p_{T}^{2}}{2m^{2}\omega_{c}^{2}}\sin ^{2}%
\frac{\omega_{c}t}{2}\right]. 
\end{eqnarray}%
As expected, in all these 2D realizations the returns of the spin to the initial 
$S(0)-$value occur after each cyclotron period.


\begin{thebibliography}{99}
\bibitem{Stanescu} T. Stanescu, B. Anderson, and V. Galitski, Phys. Rev. A 
\textbf{78}, 023616 (2008).




\bibitem{Lin} Y.-J. Lin, K. Jim\'{e}nez-Garci\'{a}, and I. B. Spielman,
Nature \textbf{471}, 83 (2011).




\bibitem{Sommer} A. Sommer, M. Ku, G. Roati, and M. W. Zwierlein, Nature 
\textbf{472}, 201 (2011).




\bibitem{Liu} X.-J. Liu, M. F. Borunda, X. Liu, and J. Sinova, Phys. Rev.
Lett. \textbf{102}, 046402 (2009).




\bibitem{Wang} P. Wang, Z.-Q. Yu, Z. Fu, J. Miao, L. Huang, S. Chai, H.
Zhai, and J. Zhang, Phys. Rev. Lett. \textbf{109}, 095301 (2012).




\bibitem{Cheuk} L. W. Cheuk, A. T. Sommer, Z. Hadzibabic, T. Yefsah, W. S.
Bakr, and M. W. Zwierlein, Phys. Rev. Lett. \textbf{109}, 095302 (2012).




\bibitem{Huang16} L. Huang, Z. Meng, P. Wang,	P. Peng, S.-L. Zhang, L. Chen,	
D. Li,	Q. Zhou, and J. Zhang, Nature Physics, in print  (2016)



\bibitem{Winkler03} R. Winkler, \textit{Spin-orbit Coupling Effects in
Two-Dimensional Electron and Hole Systems}, Springer Tracts in Modern
Physics (2003).




\bibitem{Zutic04} J. Fabian, A. Matos-Abiague, C. Ertler, P. Stano, and I.
Zutic, Acta Physica Slovaca \textbf{57}, 565 (2007).




\bibitem{Dyakonov08} \textit{Spin Physics in Semiconductors}, Springer
Series in Solid-State Sciences, Ed. by M.I. Dyakonov, Springer (2008).




\bibitem{Wu10} M.W. Wu, J.H. Jiang, and M.Q. Weng, Phys. Reports \textbf{493}, 61 (2010).




\bibitem{Lin09} Y.-J. Lin, R. L. Compton, K. Jim\'{e}nez-Garci\'{a}. J. V.
Porto, and I. B. Spielman, Nature \textbf{462}, 628 (2009).




\bibitem{Cooper} N. R. Cooper, Adv. Phys. \textbf{57}, 539 (2008).




\bibitem{reviews} J. Dalibard, F. Gerbier, G. Juzeliunas, and P. \"{O}hberg,
Rev. Mod. Phys. \textbf{83}, 1523 (2011); I. B. Spielman, Ann. Rev. Cold
At. Mol. \textbf{1}, 145 (2012); H. Zhai, Int. J. Mod. Phys. B \textbf{26}, 1230001 (2012); 
V. Galitski and I. B. Spielman, Nature \textbf{494}, 49 (2013); X. Zhou, Y. Li, Z. Cai, 
and C. Wu, J. Phys. B: At. Mol. Opt. Phys. \textbf{46}, 134001 (2013); 
N. Goldman, G. Juzeliunas, P. \"{O}hberg, and I. B. Spielman, 
Rep. Progr. Phys. \textbf{77}, 126401 (2014).




\bibitem{note1} The units with $\hbar \equiv 1$ and $c\equiv 1$ are employed
and the summation over repeated indices is assumed throughout
the paper.




\bibitem{Mineev92} V. P. Mineev and G. E. Volovik, Journal of Low
Temperature Physics \textbf{89}, 823 (1992).




\bibitem{Frolich93} J. Fr\"{o}hlich and U. M. Studer, Rev. Mod Phys. \textbf{65}, 733 (1993).




\bibitem{Aleiner01} I. L. Aleiner and V. I. Fal'ko, Phys. Rev. Lett. \textbf{87}, 256801 (2001).




\bibitem{Levitov03} L. S. Levitov and E. I. Rashba, Phys. Rev. B \textbf{67}, 115324 (2003).




\bibitem{Tokatly08} I. V. Tokatly, Phys. Rev. Lett. \textbf{101}, 106601
(2008).




\bibitem{Leurs08} B. W. A. Leurs, Z. Nazario, D.I. Santiago, and J. Zaanen,
Annals of Physics \textbf{323}, 907 (2008).




\bibitem{Raimondi} R. Raimondi and P. Schwab, Europhys. Lett. \textbf{87},
37008 (2009); M. Milletar, R. Raimondi, and P. Schwab, Europhys. Lett. 
\textbf{82}, 67005 (2008).




\bibitem{Bernevig06} B. A. Bernevig, J. Orenstein, and S.-C. Zhang, Phys.
Rev. Lett. \textbf{97}, 236601 (2006).




\bibitem{Koralek09} J. D. Koralek, C. P. Weber, J. Orenstein, B. A.
Bernevig, Shou-Cheng Zhang, S. Mack, and D. D. Awschalom, Nature \textbf{458}, 610 (2009).



\bibitem{Schliemann16} A review on the spin-helix related effects
in semiconductors can be found in: J. Schliemann, arXiv:1604.02026.



\bibitem{Liu09} For calculation of synthetic spin-orbit  coupling 
in cold Fermi gases see: X.-J. Liu, M. F. Borunda, X. Liu, and J. Sinova, Phys. Rev. Lett. 
\textbf{102}, 046402 (2009).



\bibitem{Korn} T. Korn, Phys. Reports, \textbf{494}, 415 (2010). 



\bibitem{Tokatly10a} I. V. Tokatly and E.Ya. Sherman, Annals of Physics 
\textbf{325}, 1104 (2010) 




\bibitem{Tokatly10b} I. V. Tokatly and E.Ya. Sherman, Phys. Rev. B 
\textbf{82}, 161305 (2010).


\bibitem{Kirzhnitz} D.A Kirzhnitz, Yu.E. Lozovik, and G.V. Shpatakovskaya,
Soviet Physics - Uspekhi \textbf{18}, 649 (1977).




\bibitem{small_t_bosons} The small-$\xi(t)$ behavior obtained with the exact Bose-distribution
is parabolic in time as a consequence of the $f-$sum rule \cite{Tokatly10b,Vignale}.




\bibitem{nondegenerate} For non-degenerate gases with $f(p)\sim\exp \left(
-p^{2}/2p_{T}^{2}\right)$ the function becomes $\mathcal{N}({\xi }(t))\sim\exp [-p_{T}^{2}{\xi }^{2}(t)/2].$




\bibitem{Mermin} N. D. Mermin, Phys. Rev. B \textbf{1}, 2362 (1970).



\bibitem{Tokatly13} I. V. Tokatly and E.Ya. Sherman, Phys. Rev. A 
\textbf{87}, 041602 (2013).




\bibitem{Forster} D. Forster, \textit{Hydrodynamic Fluctuations, Broken
Symmetry, and Correlation Functions} (Addison-Wesley Publishing, Reading,
Masachusetts, 1994).

\bibitem{TYu} Spin diffusion in ultracold cold Fermi-gases in the presence of the spin-orbit coupling and
Zeeman fields only has been studied by T. Yu and M. W. Wu, Phys. Rev. A \textbf{92}, 013607 (2015)
by using the kinetic spin Bloch equations.


\bibitem{Vignale} G. Giuliani and G. Vignale, \textit{Quantum Theory of the
Electron Liquid}, Cambridge University Press (2005).



\bibitem{Dyakonov73} M. I. Dyakonov and V. I. Perel', Sov. Phys. Solid State 
\textbf{13}, 3023 (1972).




\bibitem{Ivchenko} E. L. Ivchenko, Fiz. Tverd. Tela (Leningrad) \textbf{15},
1566 (1973) [Sov. Phys. Solid State \textbf{15}, 1048 (1973)]




\bibitem{Burkov} A. A. Burkov and L. Balents, Phys. Rev. B \textbf{69},
245312 (2004).



\bibitem{D_bosons} Since for the Bose gas at a temperature slightly above the condensation,
both the momentum $p_{\mu}$ and the diffusion coefficient $D\sim p_{\mu}$ (at a given $\tau$) tend to zero, the spin
relaxation near the Bose-Einstein condensation transition is strongly suppressed.




\bibitem{Glazov07} M. M. Glazov, Solid State Commun. \textbf{142}, 531
(2007).













%

%

%

%

%

\end{thebibliography}
\end{document}